# Via Method for Lithography Free Contact and Preservation of 2D Materials

*Evan J. Telford[‡†], Avishai Benyamini[‡||], Daniel Rhodes[||], Da Wang[†], Younghun Jung[||], Amirali Zangiabadi[⌐], Kenji Watanabe[§], Takashi Taniguchi[§], Shuang Jia[||⊥], Katayun Barmak[⌐], Abhay N. Pasupathy[†], Cory R. Dean[†], James Hone\*[†]*

† Department of Physics, Columbia University, New York, NY, USA

‡ Department of Mechanical Engineering, Columbia University, New York, NY, USA

⌐Department of Applied Physics and Applied Mathematics, Columbia University, New York, NY, USA

§ National Institute for Materials Science, 1-1 Namiki, Tsukuba, 305-0044 Japan

|| International Center for Quantum Materials, School of Physics, Peking University, Beijing, China

⊥ Collaborative Innovation Center of Quantum Matter, Beijing, 100871, China

‡ These authors contributed equally to this work

ABSTRACT:

Atomically thin 2D materials span the common components of electronic circuits as metals, semi-conductors, and insulators, and can manifest correlated phases such as superconductivity, charge density waves, and magnetism. An ongoing challenge in the field is to incorporate these 2D materials into multi-layer hetero-structures with robust electrical contacts while preventing disorder and degradation. In particular, preserving and studying air-sensitive 2D materials has presented a significant challenge since they readily oxidize under atmospheric conditions. We report a new technique for contacting 2D materials, in which metal *via contacts* are integrated into flakes of insulating hexagonal boron nitride, and then placed onto the desired conducting 2D layer, avoiding direct lithographic patterning onto the 2D conductor. The metal contacts are planar with the bottom surface of the boron nitride and form robust contacts to multiple 2D materials. These structures protect air-sensitive 2D materials for months with no degradation in performance. This via contact technique will provide the capability to produce 'atomic printed circuit boards' that can form the basis of more complex multi-layer heterostructures.

TEXT:

The variety of electronic properties exhibited by two-dimensional (2D) materials is promising for the realization of versatile atomic scale circuits and multi-functional devices[1-3]. These materials include metals, semiconductors, and large band gap insulators that can be combined to create versatile atomic scale resistors, capacitors, transistors, and other circuit elements[12-14]. Moreover, many of the 2D materials exhibit more exotic properties including charge density waves,[4-5] superconductivity,[6-8] and magnetism[9-11]. One of the main challenges in the study of 2D materials has been to achieve robust electrical contacts, and recent work has demonstrated a number of techniques toward this goal[15-24]. However, further advances are still needed. For instance, residue and structural damage from lithographic patterning of contacts can induce tunnel barriers and Fermi level pinning, both of which can increase contact resistance. This motivates exploration of techniques that avoid direct lithographic processing on the material under investigation. Moreover, most of the metallic 2D materials as well as small-gap semiconductors oxidize on relatively short time scales under atmospheric conditions[25-26] altering their electronic characteristics in undesirable and uncontrolled ways. Creating robust, stable contacts to these materials still presents a significant challenge. While hexagonal boron nitride (hBN) can act as an excellent encapsulant, contact areas must necessarily be exposed[23]. Therefore, approaches employing graphene or pre-patterned metal contacts in conjunction with hBN have been pursued[4-8,27]. These techniques allow fabrication of high-quality devices, but the device contacts tend to suffer from reliability issues, likely because the hBN cannot make a perfect seal around the protruding graphene or metal contact. Moreover, pre-patterned metal contacts have ~1-2 nm roughness[30] (compared to atomically flat 2D materials) due to the intrinsic grain size of the metals used, and are not planar with the substrate, creating local strains and non-uniform contact to the material, both of which can impact the overall quality of the electronic contact.[31]

Here we report a novel technique in which metal contacts are first embedded within hBN, and then laminated onto the desired 2D material to simultaneously achieve electrical contact and encapsulation without direct lithographic patterning. The embedded metal provides an electrical connection between the contacted material and the top of the hBN layer, analogous to the 'vias' in printed circuit boards (PCBs). We find that the via contacts are highly flat and planar with the hBN surface, and make intimate contact to the hBN sidewalls. The contacts are of high quality for a variety of 2D materials, and air-sensitive materials show superior stability, with no change in behavior over many months.

Figure 1a shows a schematic of the via contact fabrication and assembly process. hBN flakes (typically 20-30 nm thick) are exfoliated onto 285nm $SiO_2$ substrate and identified by optical contrast[28-29]. Next, electron beam lithography and reactive ion etching (RIE) is used to etch holes through the hBN to create the vias. For this work, etching was performed in an Oxford Plasmalab 100 ICP-RIE using either $SF_6$:$O_2$ or $CHF_3$:$O_2$ 40:4 sccm gas mixture with 60W RF power and 40mTorr gas pressure. We note that $SF_6$ is more selective than $CHF_3$ but leaves the $SiO_2$ surface somewhat rougher (120 pm for $CHF_3$ and 145 pm for $SF_6$). A second lithography step followed by electron beam evaporation is then used to pattern metallic contacts that fill the holes and extend slightly beyond to the top hBN surface. Noble metals that do not adhere well to $SiO_2$ (including Au, Pd, and Pt) are used without the usual adhesion layer. The result is an hBN flake with an array of embedded metallic contacts, which can then be picked up and transferred onto additional 2D layers using the dry polymer transfer technique[18] (Fig. 1b), simultaneously encapsulating and electrically contacting the 2D material. Finally, standard lithographic techniques are used to connect the vias to bonding pads. A false-color scanning electron microscope (SEM) image of a finished device is shown in figure 1c.

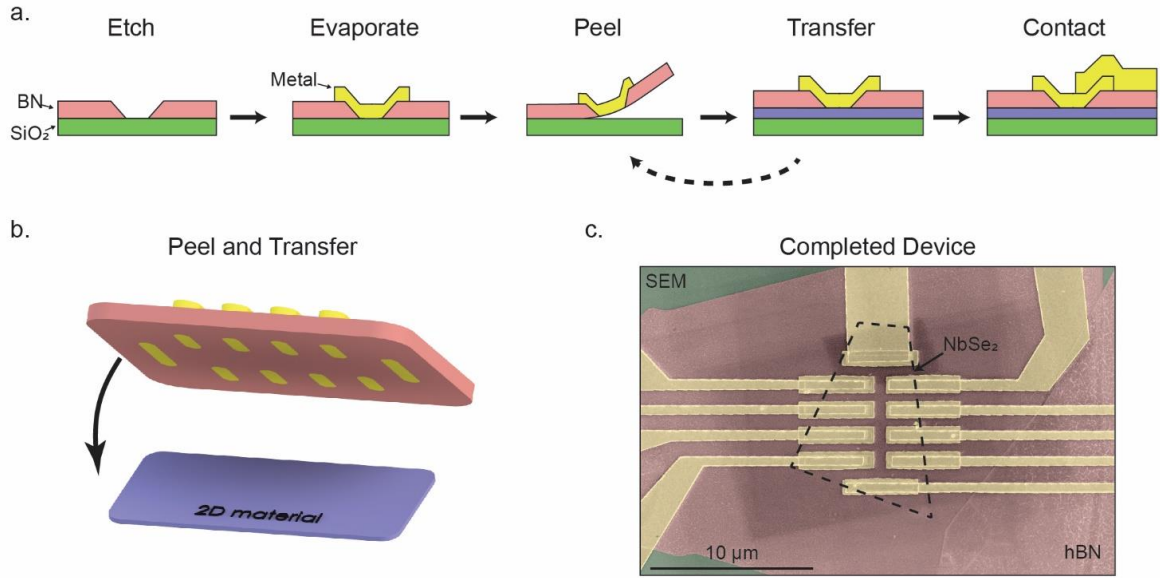

**Figure 1.** Process flow for via contacts and device assembly. (**a**) Side view diagrams of the via contact fabrication process and application in contacting atomic scale flakes. (b) An angle view diagram of placing a via contact on top of a desired 2D material. (c) False-color SEM image of a via contact transferred on top of a monolayer of NbSe$_2$. The edge of the NbSe$_2$ flake is outlined by the black dashed line. Color code: yellow, Au; green, SiO$_2$; red, hBN.

Figure 2 examines the physical structure of the via contact. Figure 2a shows a false-color transmission electron microscope (TEM) image of a cross-section of a single gold via contact on monolayer 2H-NbSe$_2$. The structure of the via corresponds well to the schematic shown in Figure 1, with the metal sealed to the hBN sidewall and planar with the bottom of the hBN. Figure 2b examines these features at greater resolution by high resolution TEM (HRTEM) imaging. The metal is flush with the hBN except for a small void at the bottom, demonstrating how the embedded metallic contacts maintain the encapsulation capabilities of the hBN layer. The void, likely due to the surface tension of the metal or a plastic deformation when the via is repeatedly picked up and placed down, extends ~12 nm from the edge of the hBN, after which the metal is planar with the hBN on the scale of the substrate roughness. The size of the void is sample dependent and on the scale of 10-20 nm. The 2D material deforms slightly to adhere to the metal in the void region; future work will examine whether this deformation adversely affects contacts and strategies to minimize this effect. The composition of the contact area is confirmed by energy dispersive spectroscopy (EDS) on the cross-sectional cut, distinguishing the elemental composition of the air sensitive flake and the via contact layer (Fig. 2c).

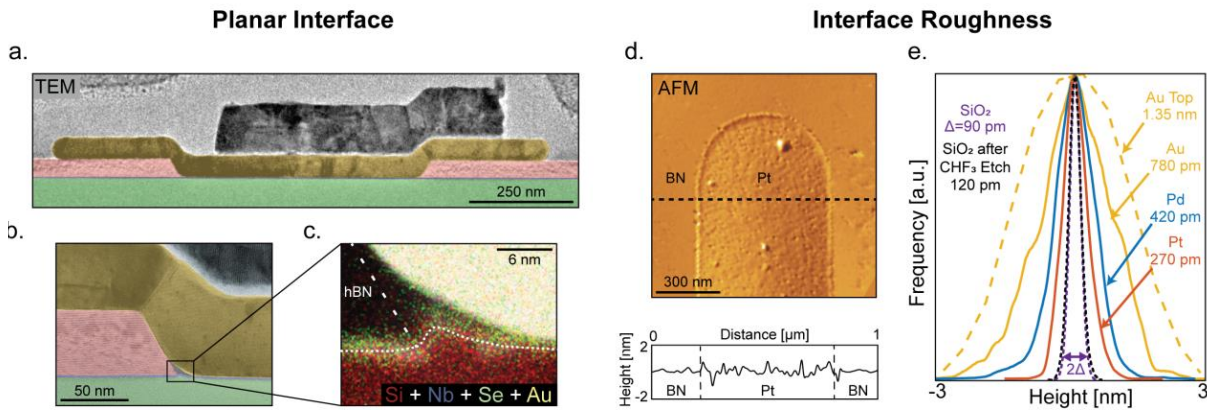

**Figure 2.** Planarity and smoothness of via contact interface. **(a)** False-color side view TEM image of a cross section from a gold via contact on monolayer NbSe₂. Red, hBN; green, SiO₂; yellow, Au; blue, NbSe₂. (b) HRTEM image of the edge and of the interface between the contacted NbSe₂ and the via contact. (c) An EDS (energy dispersive spectroscopy) TEM image of a close-up region shown in panel b. The elements identified are Nb, Se, Si, and Au. The Si, Nb, and Se appear present in the gold contact region due to secondary scattering effects. The white dashed and dotted lines represent the hBN-void boundary, observed in Figure 2b, and the position of the monolayer NbSe₂ respectively, both identified from HRTEM image of the same region (not shown). The monolayer NbSe₂ thickness, $\sim 0.6\,nm$, is observed thicker probably due to scattering from other atoms. The void area may contain residual materials from the FIB process. (d) Top: AFM scan of the bottom side of a single via contact. The amplitude error is shown to highlight the edges of the metal contact. Bottom: height profile cut through the dashed black line in the top AFM image. (e) Height histograms for bare and etched SiO₂, metal roughness on the SiO₂ side after pick-up and flipping (averaged over 3 contacts for each metal), and a comparison to the roughness of a deposited gold surface. The half width of the histograms is equated with the roughness of the material. Each histogram is generated from ~300nm x 300nm square AFM area.

The morphology of the bottom surface of the via contact is examined by atomic force microscope (AFM) imaging of an inverted hBN flake (Figure 2d). Consistent with the TEM cross-section imaging, we observe no step between the metallic contact and the hBN layer. Because the bottom surface of the metal is templated by the flat $SiO_2$ surface, its roughness should closely match that of the $SiO_2$. This is studied by plotting AFM height histograms of the $SiO_2$ and metal surfaces (Figure 2e). The $SiO_2$ is flat (90 pm half width) and is only slightly roughened upon etching (to 120 pm for $CHF_3$ and 145 pm for $SF_6$). Metal (Au, Pd, Pt) vias are somewhat rougher, but all are still quite flat (270, 420, and 780 pm, respectively). For comparison, the top surface of the evaporated Au has roughly twice the roughness of the templated bottom surface.

To compare the via technique to established contact and preservation techniques we assembled devices from graphene, 2H-NbSe$_2$ and β-MoTe$_2$. Figure 3a shows the electronic characteristics of via contacted bilayer graphene along with an optical image of the device. The device consists of bilayer graphene fully encapsulated by hBN on top of a graphite flake acting as an electrostatic gate. The right panel shows 2-probe and 4-probe resistances of bilayer graphene as a function of gate bias (fridge line resistance subtracted from the 2-probe measurement). Near zero bias, we observe the characteristic resistance increase around the charge neutrality point, with a disorder density broadening of $\Delta n = 6.8 \cdot 10^{10}$ cm$^{-2}$ at 1.7 K, comparable to that seen in similar edge-contacted devices[19]. At high gate biases (high graphene densities), the bilayer graphene resistance dramatically decreases, and the 2-probe resistance is dominated by the contact resistance. The data shows remarkably low contact resistance using the via contact method, $130 \; \Omega \cdot \mu m$ (determined at $n_e = -5.5 \cdot 10^{12}$ cm$^{-2}$), comparable to the best results of other contact techniques, a comparison of which is shown in figure 4a.

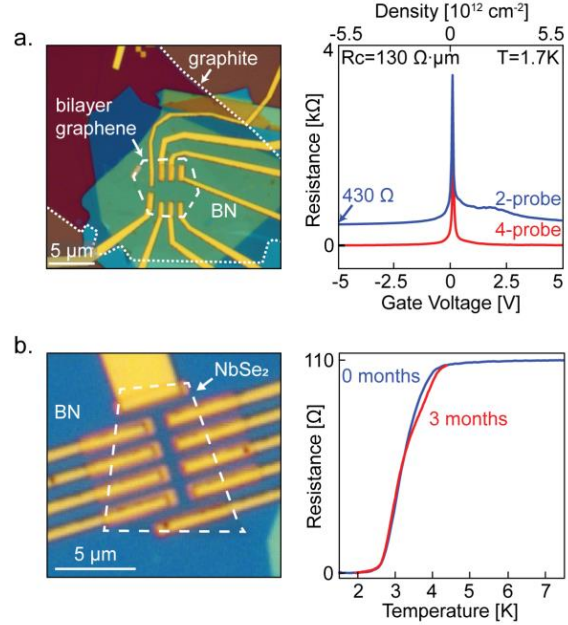

**Figure 3.** Assembled devices and measurements of different classes of low dimensional materials. **(a)** Graphene via device consisting of a bilayer graphene flake fully encapsulated with hBN on top of a graphite gate. Left: optical image of the device. Right: a measurement of bilayer graphene 2-probe and 4-probe resistances versus graphite gate bias with the fridge line resistance subtracted. At high gate biases, the 2-probe resistance is approximately twice the contact resistance of the via contacts. The noted contact resistance is extracted at a density $n_e = -5.5 \cdot 10^{12}$ cm$^{-2}$. The scaling between gate voltage and carrier density was determined from the relationship between integer quantum hall filling fractions and gate voltage given by $\frac{h}{e^2} \frac{1}{\nu} = \frac{B}{n_e e}$ (measured at 9T magnetic field). The gate capacitance $C_g = n_e \, e / V_g$ determined from this method is $C_g \cong 1.8$ fF/µm$^2$, consistent with the geometric capacitance determined from the interlayer hBN spacing. (b) Monolayer NbSe$_2$ via device. Left: optical image of the device. Right: superconducting transition at different times after fabrication.

Figure 3b shows the superconducting electronic properties of a monolayer NbSe$_2$, an air sensitive superconductor. The low normal-state resistance and the superconducting transition temperature at ~3.7 K both indicate high device quality[6,7]. These properties are virtually unchanged after 3 months over multiple thermocycles and storage both in air and in a nitrogen glovebox. This provides strong indication of the quality of the encapsulation provided by the hBN with via contacts, since the superconducting properties of few layers of $NbSe_2$ are especially sensitive to oxidization.[6,7,26]

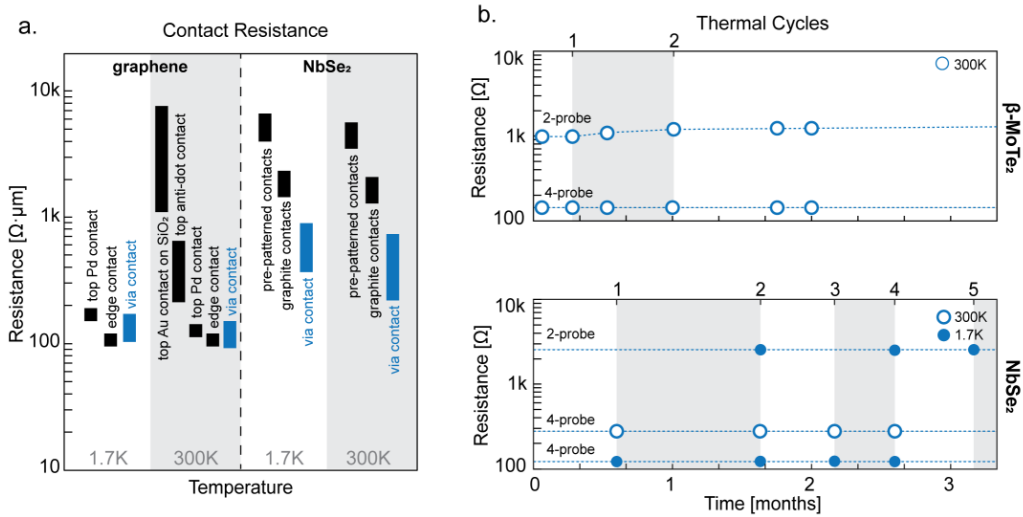

**Figure 4.** Comparison of contact resistances for different techniques and reliability of via contacts for air-sensitive materials. **(a)** Comparison for graphene to other contact techniques described in references[16-20], and for $NbSe_2$ to devices we fabricated with pre-patterned contacts (contacts deposited before $NbSe_2$ transfer) and graphite contacts (graphite used as an intermediate metallic electrode). Spread is shown over multiple measured devices or data points from references. (b) The 2-probe and 4-probe normal-state resistances of via-contacted air-sensitive 2H-$NbSe_2$ and β-$MoTe_2$ over time and thermocycles. We show here a single set of 2-probe and 4-probe measurements for each material; multiple configurations were measured and show similar stable resistances over time.

Figure 4a compares the contact resistances achieved with via contacts with those achieved for other techniques. The contact resistances of via contact devices were extracted by subtracting the material channel resistance, as determined by 4-probe resistivity measurement, and the series resistance of the probe wiring, from the measured 2-probe resistance. For graphene and $NbSe_2$ the contact resistance is scaled by the contact width.[32] We observe similar via contact resistance for graphene and bilayer graphene. The via contacts achieve contact resistances that are comparable or better than an array of already establish contact techniques[16-20] both at room temperature and at cryogenic temperatures.

To demonstrate that the technique creates robust devices from air-sensitive materials, we measured the contact and channel resistances over time and exposure to different environments for monolayer 2H-NbSe$_2$ and monolayer β-MoTe$_2$. Figure 4b shows the 2-probe and 4-probe resistances over time. The stability demonstrates the preservation of contact and sample quality, respectively. In addition to stability over time, the contacts are stable over multiple thermocycles, demonstrating their increased utility for experimental measurements.

Apart from the advantages to fundamental research, the versatility and reliability of via contacts for various types of low dimensional materials opens the road to producing PCBs on the atomic scale. The layered nature allows production of 3D circuit topologies that cannot be achieved in conventional 2D circuit architectures. The ability to create embedded contacts from any desired geometry allows for great flexibility in device design, while the ability to incorporate air sensitive materials enables incorporation of novel electronic components. Such components include superconducting materials and magnetic materials that can be used for example in vertical Josephson junctions, spin resolved transport, or in the creation of quantum-bits. The via contact method opens new possibilities to fabricate novel atomic scale circuits containing components previously inaccessible to the field of 2D nano-electronics.

AUTHOR INFORMATION

**Corresponding Authors**

*James Hone (email: jh2228@columbia.edu)

**Author Contributions**

‡ E.J.T and A.B. contributed equally to this work

**Funding Sources**

This research was primarily supported by the NSF MRSEC program through Columbia in the Center for Precision Assembly of Superstratic and Superatomic Solids (DMR-1420634), the Global Research Laboratory (GRL) Program (2016K1A1A2912707) funded by the Ministry of Science, ICT and Future Planning via the National Research Foundation of Korea (NRF), and Honda Research Institute USA Inc.

**Notes**

The authors declare no competing financial interest.

ACKNOWLEDGMENTS

A portion of this work was performed at the National High Magnetic Field Laboratory, which is supported by National Science Foundation Cooperative Agreement No. DMR-1157490 and the State of Florida. This work was performed in part at the Advanced Science Research Center NanoFabrication Facility of the Graduate Center at the City University of New York.